\documentclass[conference]{IEEEtran}
\IEEEoverridecommandlockouts
\usepackage{cite}
\usepackage{amsmath,amssymb,amsfonts}
\usepackage{graphicx}
\usepackage{textcomp}
\usepackage{xcolor}
\usepackage{subcaption}
\usepackage{bm}
\usepackage{multirow}
\usepackage{booktabs}
\usepackage{subcaption}
\usepackage{amsthm}\usepackage{balance}\usepackage{url}
\usepackage{algorithm,algpseudocode}

\def\BibTeX{{\rm B\kern-.05em{\sc i\kern-.025em b}\kern-.08em
    T\kern-.1667em\lower.7ex\hbox{E}\kern-.125emX}}

\newtheorem*{proof*}{Proof}

\algrenewcommand\algorithmicrequire{\textbf{Initialization:}}
\algrenewcommand\algorithmicensure{\textbf{Output:}}
\algdef{SE}[DOWHILE]{Do}{doWhile}{\algorithmicdo}[1]{\algorithmicwhile\ #1}%
\begin{document}

\title{Enhancement or Super-Resolution: Learning-based Adaptive Video Streaming with Client-Side Video Processing\\
\thanks{This work was supported in part by the National Natural Science
Foundation of China under Project 62101336 and in part by the Tencent Rhinoceros BirdsScientific Research Foundation for Young Teachers of Shenzhen University. J.~Yang and S.~Wang are also with the Guangdong Province Engineering Laboratory for Digital Creative Technology and the Shenzhen Key Laboratory of Digital Creative Technology. }
}

\author{Junyan Yang$^\dagger$, Yang Jiang$^\ddag$, and Shuoyao Wang$^{\dagger *}$ \thanks{$*$
Corresponding author.}\\
$^\dagger$College of Electronic and Information Engineering, Shenzhen University, China\\
$^\ddag$ Institute of High-Performance Computing, Singapore\\
E-mail: yangjunyan2020@email.szu.edu.cn, Jiang\underline{ }Yang@ihpc.a-star.edu.sg, sywang@szu.edu.cn
}


\maketitle

\begin{abstract}
The rapid development of multimedia and communication technology has resulted in an urgent need for high-quality video streaming. However, robust video streaming under fluctuating network conditions and heterogeneous client computing capabilities remains a challenge.
In this paper, we consider an enhancement-enabled video streaming network under a time-varying wireless network and limited computation capacity.  ``Enhancement'' means that the client can improve the quality of the downloaded video segments via image processing modules. We aim to design a joint bitrate adaptation and client-side enhancement algorithm toward maximizing the quality of experience (QoE).
We formulate the problem as a  Markov decision process (MDP) and propose a deep reinforcement learning (DRL)-based framework, named ENAVS. 
As video streaming quality is mainly affected by video compression, we demonstrate that the video enhancement algorithm outperforms the super-resolution algorithm in terms of signal-to-noise ratio and frames per second, suggesting a better solution for client processing in video streaming. 
Ultimately, we implement ENAVS and  demonstrate extensive testbed results under real-world bandwidth traces and videos. The simulation shows that ENAVS is capable of delivering $5\%-14\%$ more QoE under the same bandwidth and computing power conditions as conventional ABR streaming.
\end{abstract}


\section{Introduction}
Recent years have witnessed the sky-rocketing increasing network traffic. Driven by the popularity of smart user devices and communication technologies, monthly mobile data traffic is expected to exceed  396 exabytes per month by 2022, up from 122 exabytes per month in 2017. Video traffic will quadruple by 2022 and  make up an even larger percentage of total traffic than before—up to 82 percent from 75 percent \cite{Cisco}.
Video service providers, such as Youtube, Bilibili, Netflix, launch adaptive bitrate (ABR) video streaming \cite{7976298} to assure desirable user's \emph{quality of experience} (QoE)\cite{yin2015control}. 

The Dynamic Adaptive Streaming over HTTP (DASH)\cite{stockhammer2011dynamic} been recognized as a prominent video quality adaptation protocol.
In the past decade, many research efforts \cite{7485902}-\nocite{9580665}\nocite{mao2017neural}\cite{8737361} have been devoted to improve the QoE of DASH-based system.
Existing approaches are broadly classified into two categories: heuristic-based strategies and reinforcement learning (RL)-based strategies.
For instance, 
\cite{7485902} and \cite{9580665} carefully designed a heuristic rule and an online algorithm to select bitrates, respectively. 
Alternatively, \cite{mao2017neural} and \cite{8737361} investigated RL algorithms to optimize QoE by integrating with the video streaming networks. 
Such approaches, however, have failed to address the following challenges in mobile video streaming: 
\begin{itemize}
    \item \textbf{Bandwidth dependence:}
    The conventional ABR strategies highly rely on bandwidth conditions. Accordingly, the dynamic or poor bandwidth condition would mislead the prediction of bandwidth and thus degrade the QoE.
    \item \textbf{Missed effort at the client-side:}
    Nowadays, the computing power of mobile devices has skyrocketed. With client computing, video quality enhancement techniques can significantly improve the quality of videos, missed in most of the existing ABR strategies.
\end{itemize}\par
Most recently, client video super-resolution has been employed to improve the quality of the video, such as NAS \cite{yeo2018neural} and SRAVS \cite{9155384}.
Specifically, \cite{yeo2018neural} demonstrated that video super-resolution techniques successfully improve the video quality at the client-side, conditioning on a high computation capacity.
Applying real-time super-resolution in video streaming, however, involves solving a number of new and non-trivial challenges:
i) due to the high complexity of the super-resolution algorithm and strict latency requirement,  the access rate of the super-resolution module is limited, especially for the poor-computation devices.
ii) as shown in the preliminary experiments in Section II,  the quality loss in video streaming is more significant by the MPEG/JPEG compression rather than the resolution reduction under the same bitrate. 

In this paper, we propose a video enhancement based ABR video streaming framework (ENAVS). In particular, we employ a video enhancement deep learning (DL) model to enhance video quality at the client-side and deploys a deep RL (DRL) model to determine the bitrate and binary enhancement decisions (feed into the DL model or directly cache in the playback).
The main
contributions of the paper are:
\begin{itemize}
	\item To the authors’ best knowledge, this paper is the first work
that develops an online method for joint bitrate adaptation and client-side video enhancement. 
\item We formulate the sequential decision-making problem as a Markov decision process (MDP) problem and propose a DRL framework based on the actor-critic learning framework. To speed up the convergence,  the entropy term and predicted re-buffering time are introduced to the objective and the reward, respectively.
	\item We carry out extensive evaluations over real-world bandwidth traces, where ENAVS provides  $5\%-14\%$ extra QoE under the same bandwidth and computing capability.
\end{itemize}

\section{Preliminary}
Recently, the client processing algorithms have shown great potential to improve the quality of video streaming.
{However, under a same bitrate, the quality loss of resolution reduction is greater than that of compression, and thus enhancement model might be a better alternative for client enchantment.}
In this section, we present the preliminary experiments and find that \emph{enhancement algorithms outperform the super-resolution algorithms in terms of both Peak Signal-to-Noise Ratio (PSNR) and Frame Per Second (FPS)} in video streaming network. 
\subsection{Experiment Setups}
Firstly, we select a series of 1K source videos from Bilibili (a streaming website from China).
Then, we use the FFMPEG application to compress the source videos into multi-bitrate videos.
We employ two image enhancement algorithms and one image super-resolution algorithm, i.e., DNCNN\cite{7839189}, CBDNet\cite{guo2019toward}, and SRCNN\cite{dong2014learning} (employed in  SRAVS\cite{9155384}), to process the compressed video frames, respectively.
The main performance metrics are the quality of the video frames and the processing speed of the three models, i.e., PSNR and FPS.
{Specifically, we feed the compressed video frames with 1k resolution and 2Mbps bitrate into the first two models, the compressed video frames with 540p resolution and 2Mbps bitrate into the last model, and the models return the processed video frames.}
All the computations are executed on a machine with an Intel i7-10700F CPU, an NVIDIA RTX 2070s GPU, and 32GB RAM.  

\subsection{Test Results}

\begin{figure}
\centering
\begin{subfigure}{.49\linewidth}
\centering
\includegraphics[width=\linewidth]{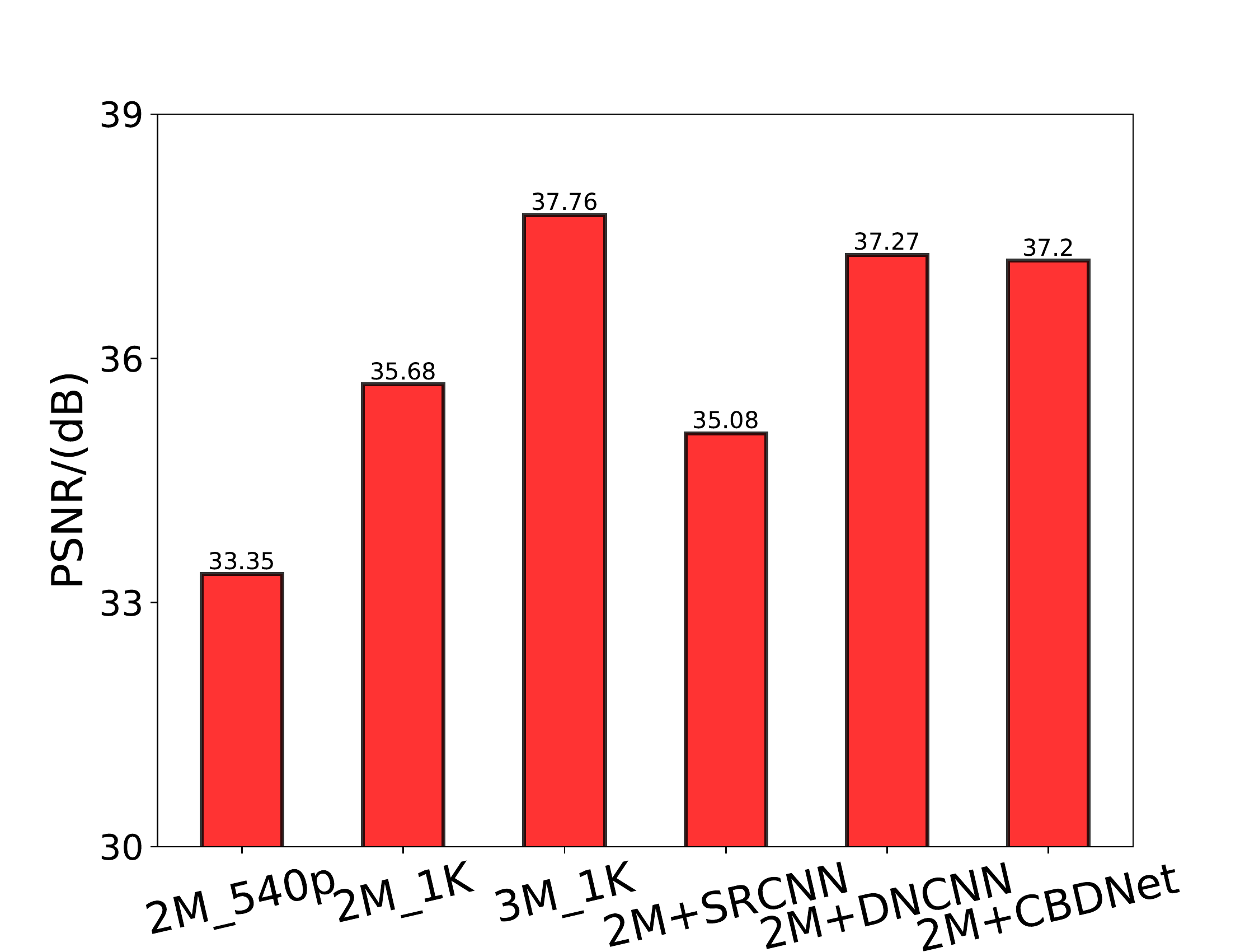}
\caption{PSNR}
\end{subfigure}
\begin{subfigure}{.49\linewidth}
\centering
\includegraphics[width=\linewidth]{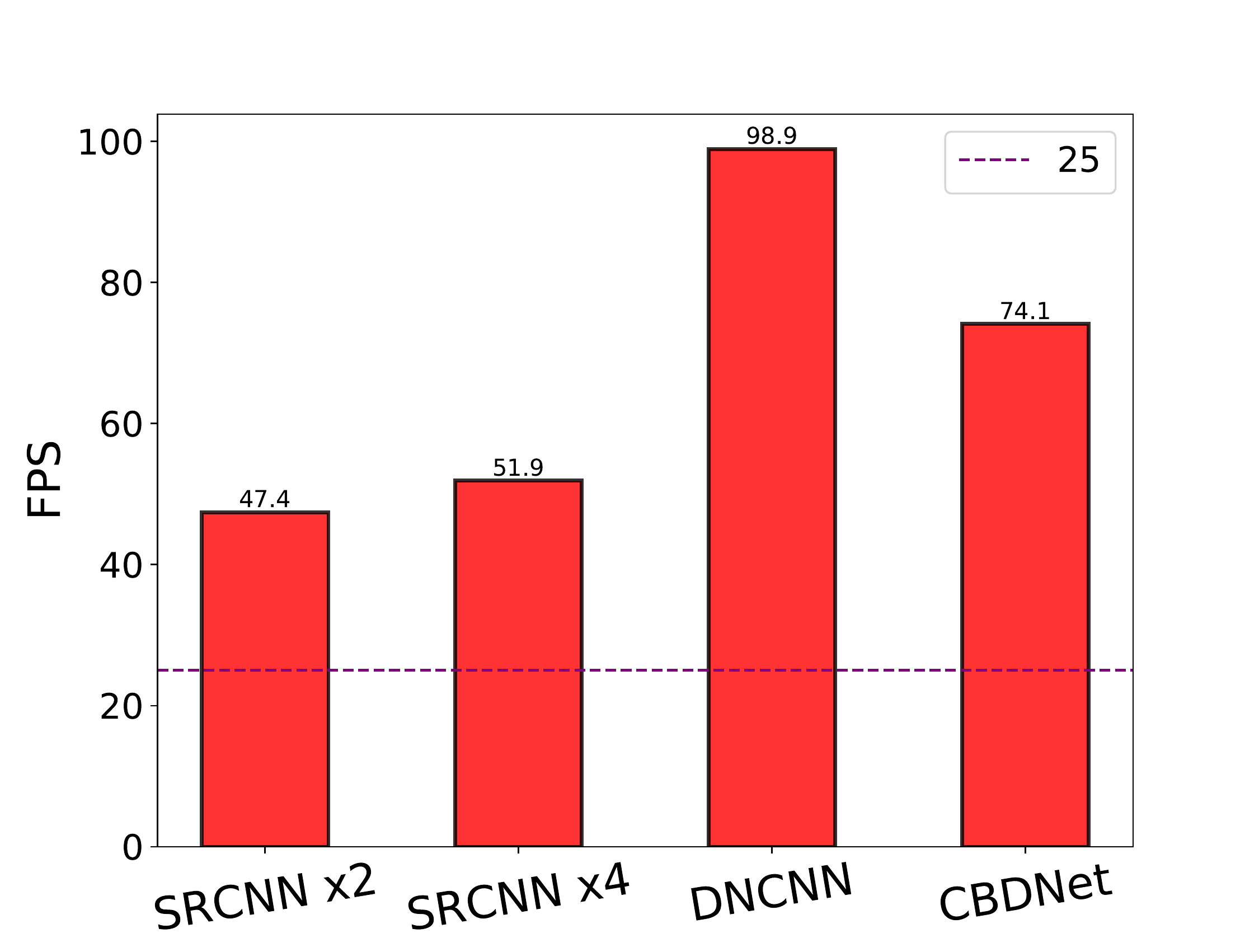}
 \caption{FPS}
\end{subfigure}
\caption{The PSNR and FPS comparison versus different client DL models.}\label{fig:gln} 
\end{figure}

{As shown in Fig.~\ref{fig:gln}a, we can find that the PSNR of 2Mbps with 540p resolution, 2Mbps with 1K resolution, 3Mbps with 1K resolution, 2Mbps with 540p resolution after SRCNN, 2Mbps with 1K resolution after DNCNN, and 2Mbps with 1K resolution after CBDNet are  $33.35$, $35.68$, $37.76$, $35.08$, $37.20$, and $37.27$, respectively.}
This indicates that the enhancement model can improve the quality of the 2Mbps video frames. Specifically,  video quality at 2Mbps is comparable to video at 3Mbps after the enhancement, i.e., $37.76$ and $37.27$.
However, the PSNR of the video returned by SRCNN is even lower than the original 2Mbps {with 1K resolution} frames. 
This validates our intuition that \emph{{the quality loss of resolution reduction is greater than that of compression, and thus enhancement model might be a better alternative for client enchantment}}.
Moreover, we plot the comparison of FPS of the three models in  Fig.~\ref{fig:gln}b. 
Since the FPS of modern video streaming is mostly 15 to 25 FPS, we also plot the red dash line of 25 FPS as a reference.
We can observe from Fig.~\ref{fig:gln}b  that computation latency of DNCNN is much smaller than other two models.
Specifically, the FPS of SRCNNx2, SRCNNx4, DNCNN, and CBDnet are $47.4$, $51.9$, $98.9$, and $74.1$, where  SRCNN is only half of DNCNN.
\footnote{Here, x2 and x4 are the upscaling factor of SRCNN.}


\subsection{Summary}
Our experiments reveal that enhancement models could be a better alternative of super-resolution models in ABR streaming.
In particular, the PSNR of video at 2Mbps is comparable to 3Mbps after the enhancement. 
Inspired by the preliminary experiments, we investigate the joint bitrate adaptation and client enhancement problem for wireless video streaming networks, to fully utilize the client computation capability.

\section{System Overview}
\subsection{Adaptive Video Streaming}
Particularly, we consider a most typical ABR video streaming protocol in the video-on-demand (VoD) streaming system. Its primary features are summarized as follows.\par
\subsubsection{Video Chunks}
In the VoD streaming system, a video is compressed as a series of images.
An intact video file is divided into a sequence of chunks with a fixed playback time  $T$. 
The chunks are indexed from 1 to $N$, where $N$ represents the number of chunks of the source video. \par
\subsubsection{Multi-Bitrate}
A chunk is encoded into multiple bitrate copies on the server-side. As a result, users can choose the most appropriate bitrate for each chunk based on their own playback statistics and the network conditions.
For notation simplicity, we denote by $\mathcal{R}$ and $R_i \in \mathcal{R}$ as the bitrate set and the requested bitrate of the $i$-th chunk, respectively.
\subsection{Network Model}
\begin{figure}[t] 
\centering
\includegraphics[width=\linewidth]{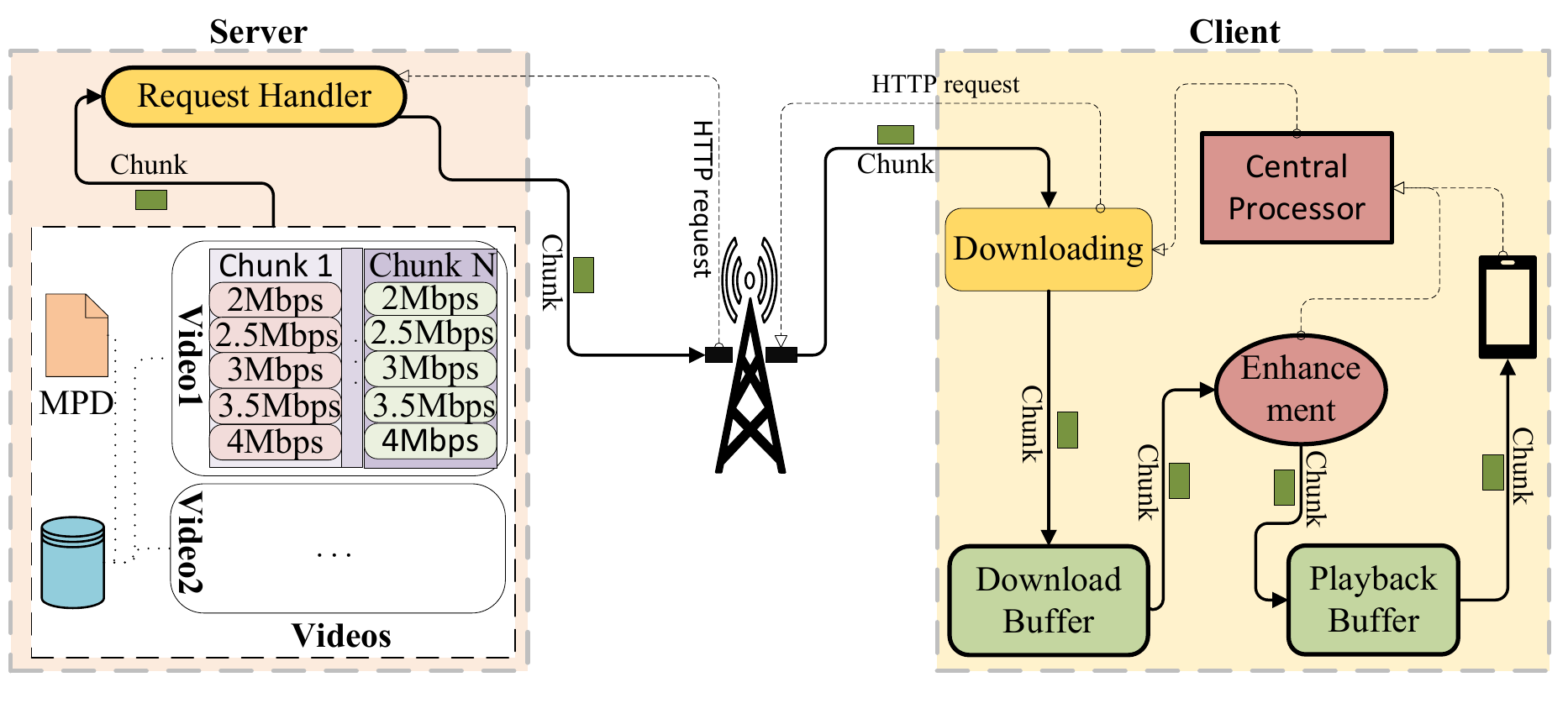}
\caption{The overview of system}\label{system}
\end{figure}
As shown in Fig.~\ref{system}, we consider a wireless video streaming network with client-side video processing.
The network consists of a mobile client and a video server. 
The mobile client involves a central processor controller, a download buffer (DB), an enhancement module (EN), a playback buffer (PB), and a player.  
The content server caches video data as multi-bitrate chunks and a media presentation description (MPD) file. 
The client sends a request signal to the server through the wireless uplink, and the server transmits the requested video chunks of the requested bitrate through the wireless downlink.
\subsection{Video Chunks with Client Processing}
In ENAVS, client-side processing is introduced into adaptive video streaming. 
The decisions in the processing-enabled video streaming network are: i) the chunk bitrate, and ii) whether to call the client processing.
Without loss of generality, we denote by $\mathcal{A}$, $a_i=(R_i,p_i) \in \mathcal{A}$, and  $p_i \in (0,1)$ as the set of all feasible decisions, the decision on chunk $i$, and the enhancement factor of chunk $i$, respectively.
Here, $p_i = 1$ means the client feeds the chunk $i$ into the DL model and $p_i = 0$ means the client directly caches the chunk $i$ in the playback buffer. 
To determine the quality of the generated video chunk, we measure the quality of the final video chunk by PSNR.
We define PSNR($a_i$) as the PSNR of the $i$-th chunk under a legal decision $a_i$. The PSNR map $\{\text{PSNR}(a_i)\}_{a_i \in \mathcal{A}}$ is pre-calculated and recorded in the MPD file.

In ENAVS, the enhancement algorithm DNCNN \cite{7839189} is deployed on the mobile client as the enhancement module, which enhances the quality of the downloaded low-bitrate video chunk.
DNCNN is constructed based on 2D-convolution layers and adopted the residual learning strategy\cite{7780459}.
DNCNN  takes a low-bitrate video frame as input and improves the quality of the input frame.

\section{Adaptive Video Streaming with video enhancement}
In this section, we formulate the problem of bitrate adaptation with video enhancement, and propose the DRL-based adaptive video streaming policy, named ENAVS. 
\subsection{Timeline}
There are two buffers in the proposed ENAVS.
During the playback, the mobile client downloads the requested chunks into the download buffer.
Simultaneously, the enhancement module extracts the full downloaded chunks from the download buffer and push the chunks to the playback buffer after the enhancement is finished.
Without loss of generality, we denote by $B^D_{i}$ and $B^P_{i}$ as the lengths of the downloading buffer and the playback buffer when the chunk $i$ is newly cached, respectively.
Due to the physical limitation of the mobile memory, the lengths of the downloading buffer and the playback buffer are upper bounded by $\overline{B^D}$ and $\overline{B^P}$, respectively. The timeline of each chunk is described as follows.

\subsubsection{Video Downloading}
During the playback, the mobile client downloads the requested chunks into the download buffer through a wireless downlink.
The downloading time $\tau^D_i$ of the $i$-th chunk is
\begin{equation}
     \tau^D_i = \frac{R_i \cdot  T}{c_i},
\end{equation}
where $c_i$ denotes the average bandwidth during the download process of the $i$-th chunk.
If the downloading buffer is not full $B^D_{i}<\overline{B^D}$,  the client immediately sends the pull request with the selected bitrate for the $i$-th chunk to the server. Otherwise, the client keeps waiting  until the $(i-\overline{B^D})$-th chunk enters the processing module and releases the buffer occupancy. 
That is, 
\begin{equation}
    t_i^D =\begin{cases}
    0&  \text{$i=1$}\\
     \max (t_{i-1}^D +  \tau^D_{i-1} , t^E_{i-\overline{B^D}})& \text{otherwise},
    \end{cases} 
\end{equation}
where $t_i^D$ and $t^E_i$ denote the request and enhancement starting time for chunk $i$, respectively.

\subsubsection{Video Processing}
When the $i$-th chunk is cached in the downloading buffer, the client could employ the client-side DL models to process the chunk and place the processed chunk into the playback buffer, where the processing time $\tau^E_i>0$.
Alternatively, the client could also directly place the cached chunk into the playback buffer, i.e., $\tau^E_i=0$.
The enhancement starting time can be expressed as 
\begin{equation}
    t_i^E =\begin{cases}
   t_i^D + \tau^D_i&  \text{$i=1$}\\
     \max(t_{i-1}^E + \tau^E_{i-1},t_i^D + \tau^D_i,t^P_{i-\overline{B^P}})& \text{otherwise}.
    \end{cases}  
\end{equation}
Notation $t_{i-1}^E + \tau^E_{i-1}$ ensures that the DL model processes the chunks one-by-one. $t_i^D+\tau^D_i$ ensures that the enhancement is deployed after the chunk is completely downloaded. $t^P_{i-\overline{B^P}}$ ensures that the playback buffer has enough space to cache the processed chunk.

\subsubsection{Playback}
When the chunk $i$ is cached in the playback buffer and the $(i-1)$-th chunk is completed played, the client starts to play the chunk $i$. That is, the playback starting time of the $i$-th chunk $t_i^P$ can be calculated as
\begin{equation}
    t_i^P = \begin{cases}
   t_i^E + \tau_i^E&  \text{$i=1$}\\
    \max(t_{i-1}^P + T , t_i^E + \tau_i^E)& \text{otherwise}.
    \end{cases}   
\end{equation}
In particular, when $t_{i-1}^P + T<t_i^E + \tau_i^E$, re-buffering happens.
Therefore, the re-buffering time $\tau_i^r$ can be depicted by:
\begin{equation}
    \tau_i^r = t_i^P - (t_{i-1}^P + T).
\end{equation}
Overall, the downloading and playback buffers are updated as
\begin{equation}
    B^D_{i} =[i-1- \arg \underset{j}{\max} (t_j^E < t_i^D) ],
\end{equation}
\begin{equation}
     B^P_{i} =[\arg \underset{j}{\max}(t_j^E < t_i^D)- 1- \arg \underset{j}{\max} (t_j^P < t_i^D)].
\end{equation}
\subsection{QoE Objective}
In line with many existing works \cite{9155384}, we assume the QoE consist of three terms: 1) average video quality,  2) average quality variations, and  3) average re-buffering.

\subsubsection{Average Video Quality}
The traditional correlation between bitrate and chunk quality is not valid anymore due to the opportunity of video processing modules to enhance low-quality chunks. Hence, PSNR is used to quantify quality  \cite{9155384}:
\begin{equation}
\frac{1}{N} \sum_{i=1}^N \text{PSNR}(a_i).
\end{equation}
\subsubsection{Average Quality Variations}
Generally, users prefer smooth changes in video bitrates between consecutive segments. The degradation loss is defined as the sum of the magnitude changes in bitrate between consecutive segments:
\begin{equation}\label{q2}
  \frac{1}{N-1} \sum_{i=2}^N |R_i - R_{i-1}|.
\end{equation}
\subsubsection{Average Re-buffering}
Re-buffering occurs when the playback buffer is empty, and it may reduce the user's experience.
The average re-buffering time can be measured by:
\begin{equation}
  \frac{1}{N} \sum_{i=1}^N \tau_i^r.
\end{equation}

\subsection{Problem Formulation}
\begin{figure}[t] 
\centering
\includegraphics[width=\linewidth]{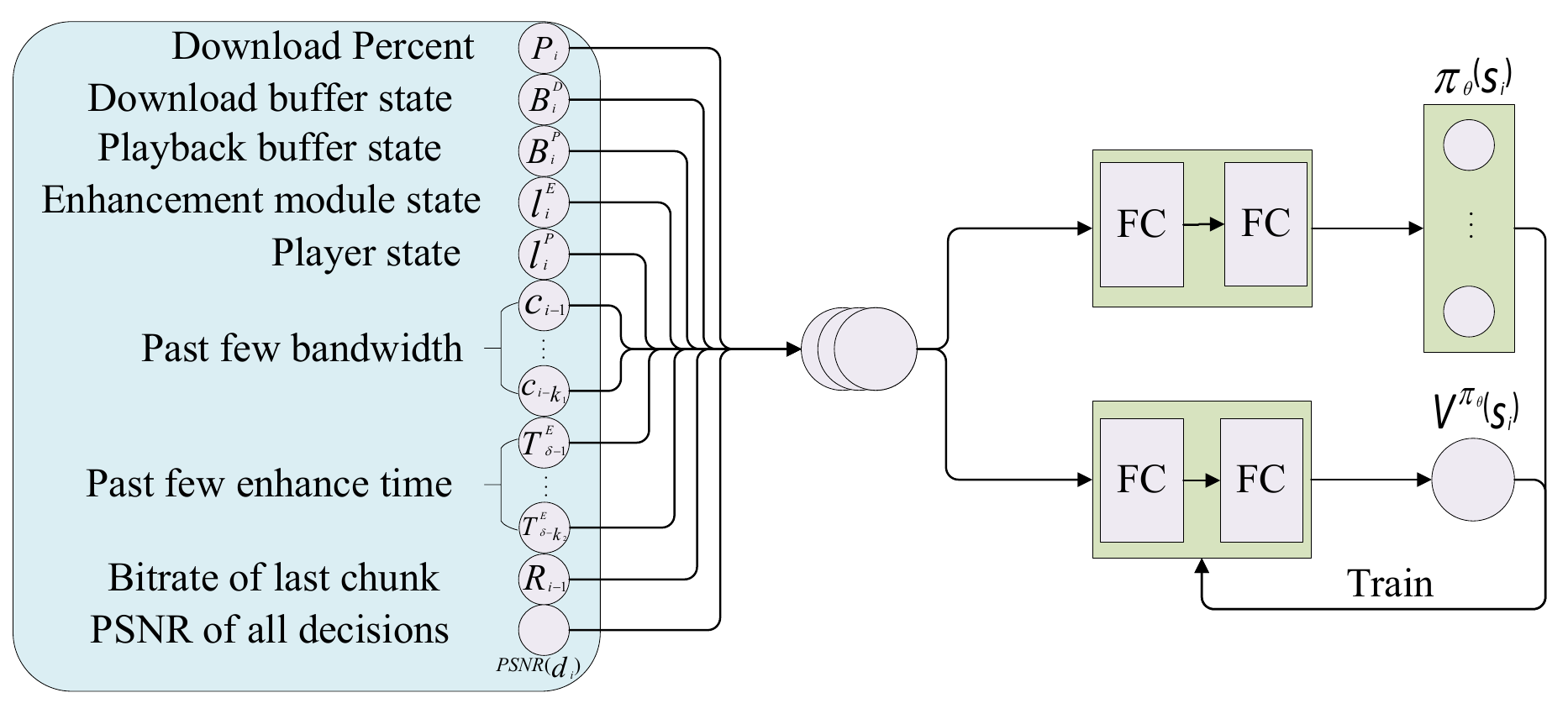}
\caption{The architecture of ENAVS}
\end{figure}
Since the unknown future bandwidth and decision coupling from (\ref{q2}), the problem is naturally an MDP, which is defined by state, action, and reward.  
In the following, we discuss the state, action as well as reward, and  design a DRL-based ABR video streaming algorithm to solve the formulated MDP problem.

\subsubsection{States}
The state at the decision of chunk $i$  is
\begin{equation}
\begin{split}
    s_i = <B^D_{i}, B^P_{i},P_i,  l_i^E, l_i^P, c_{i-1}, c_{i-2},...,c_{i-k_1}, \\
    \tau^E_{\delta}, \tau^E_{\delta -1},...,\tau^E_{\delta -k_2}, R_{i-1}, \{\text{PSNR}(a_i)\}_{a_i \in \mathcal{A}}>,
\end{split}
\end{equation} 
where $R_{i}=0, \forall i \le 0$, $\delta$ is the number of enhanced chunks.
The primary information of the system is the two buffers, i.e.,  $B^D_{i}$ and  $ B^P_{i}$. 
Moreover, the download percent of the video $P_i$ directly affects the future buffer updating.
We also consider the elapsed time of the chunk being enhance $l_i^E$ and the remaining time of the video chunk being played $l_i^P$:
\begin{equation}
    l_i^E =
    \begin{cases}
    0&  \text{$i - B^D_i = \delta + 1$}\\
    t_i^D - t_{\delta + 1}^E& \text{otherwise},
    \end{cases}
\end{equation}
and
\begin{equation}
    l_i^P = \max[T-(t_i^D - t^P_{\delta - B^P_i}),0].
\end{equation}
To cope with the time-varying network condition, we also include the average network throughput when downloading the past few chunks $c_{i-1}, c_{i-2},...,c_{i-k_1}$.
To cope with the client computation capability, we take the enhance time of the last few chunks $\tau^E_{\delta}, \tau^E_{\delta -1},...,\tau^E_{\delta -k_2}$ into consideration.
\footnote{Here, $k_1$ and $k_2$ are the historical length.}
Lastly, the bitrate of the last chunk $R_{i-1}$ and the PSNR map of chunk $i$ in MPD are also included. 
\subsubsection{Action}
At $i$-th step, the agent takes action $a_i$ according to the state $s_i$, which is treated as:
\begin{equation}
    a_i\in\mathcal{A}.
\end{equation}
\subsubsection{Reward}
In a conventional DRL environment, the reward function is formulated as the instantaneous feedback, e.g., the current QoE in video streaming, the scores in games.
To speed up the convergence, we formulate the future QoE induced by the chunk $i$, rather than the current QoE, as the reward:
\begin{equation}
    r_i = \alpha_1 \cdot \text{PSNR}(a_i) - \alpha_2 \cdot |R_i - R_{i-1}| - \alpha_3 \cdot \tau^r_i,
\end{equation}
where $\bm{\alpha}=(\alpha_1,\alpha_2,\alpha_3)$ denotes the user preference among the quality, variation, and re-buffering.
\subsubsection{MDP Formulation}
Ultimately,  the network aims to find the optimal policy $\pi$ by solving the following MDP problem:
\begin{equation}
    V_{\pi}(s)=\mathbb{E}_{\pi} \biggl[ \sum^N_{i=1} r_i \gamma^i |s \biggl],
\end{equation}
where $\pi_\theta$, $\gamma \in (0,1]$ and $s$ are the policy that maps a state to an action, the discount factor that balances the instantaneous and future rewards, and the initial network state, respectively. 

\subsection{The ENAVS Algorithm}
Particularly, we employ the asynchronous advantage actor-critic (A3C)\cite{mnih2016asynchronous} architecture
to learn the policy $\pi$. As shown in Fig. 3, A3C has two neural networks: an actor network with parameters $\theta$ that aims to make actions according to states, and a critic network with parameters $\omega$ that helps actor network to train.
The actor network and the critic network share the same input as the state $s_i$. The output of the actor network is a policy $\pi_\theta$, while the output of the critic network is the estimated state value function $V^{\pi_\theta}_{\omega}(s_i)$. 
The state value function will be used to estimate the advantage function in A3C:
\begin{equation}
    A(s_i,a_i)=r_i + \gamma V^{\pi_\theta}_{\omega}(s_{i+1})-V^{\pi_\theta}_{\omega}(s_i).
\end{equation}
Then, we add the entropy of the policy $\mathcal{H}(\pi_\theta(s_i))$ at each step, and the parameter of the actor network $\theta$ will be updated as:
\begin{equation}
    \theta = \theta + \alpha^\theta \sum^{N-1}_{i=0} A(s_i,a_i) \nabla_\theta \log \pi_\theta + \eta \mathcal{H}(\pi_\theta(s_i)),
\end{equation}
where $\alpha^\theta$ is the learning rate of $\theta$ and $\eta$ is the entropy weight.\par
The critic network aims to estimate the state value function, and we can use the mean squared error loss function to evaluate the accuracy of the prediction with parameter $\omega$:
\begin{equation}
    L(\omega) = \frac{1}{2} \sum^{N-1}_{i=0}(r_i + \gamma V^{\pi_\theta}_{\omega}(s_{i+1})-V^{\pi_\theta}_{\omega}(s_i))^2.
\end{equation}
In the case, we update the critic network parameter $\omega$ as 
\begin{equation}
    \omega \gets \omega - \alpha^\omega \nabla_\omega L(\omega),
\end{equation}
where $\alpha^\omega$ is the learning rate of $\omega$.

\section{Performance Evaluation}
In this section, we conduct evaluations over real-world traces to examine the performance of ENAVS.




\subsection{Simulation Settings}
\subsubsection{Hyper-Parameters}
In the experiments, the buffer capacity of both download buffer $B_{max}^D$ and playback buffer $B_{max}^P$ are set   {as} 5\cite{9155384}. The corresponding operations are suspended if the content in the buffer reaches the capacity.
The duration of each chunk $T$ is set   {as} 1 second. 
Furthermore, the weight of entropy $\eta=0.01$, while $k_1$ and $k_2$ are both set   {as} 8 for consideration of the past bandwidth and enhance time respectively. 
The factor discount rate of future reward $\gamma=0.9$.
\subsubsection{Video Data}
We collect 40 videos in 1K resolution   {from} 60-second length to 180-second length from Bilibili.
  {Accordingly}, with the help of the FFMPEG application, we transcode the 1K videos to five   {copies of different} bitrates: 2Mbps, 2.5Mbps, 3Mbps, 3.5Mbps, and 4Mbps.
 {For each bitrate, we
split the videos into training and testing sets is to assign 1/2 videos to the former and the remaining half to the latter.
Therefore,} the set of all decisions includes:
$\mathcal{D}=$
$[(2Mbps,0), (2Mbps,1),..., (4Mbps,0), (4Mbps,1)]$.

\subsubsection{Bandwidth Trace}
We select 40 traces of 4G bandwidth with various patterns from IDLab \cite{7546928} as the bandwidth dataset.
Considering that  {the multi-user} competition in bandwidth, we  {re-scale the original bandwidth trace, where the average bandwidth of the scaled trace falls into 2Mbps to 5Mbps.} 
We use two different methods, based on the maximum value and the average value of the entire trace, to scale the original bandwidth trace to generate two bandwidth datasets, i.e., train  and test datasets. 
\subsubsection{Computing Capacity}
Considering  {the heterogeneous devices, we model the different computational capacities with four scale factors.
First, 
we feed all the videos with different bitrate into the DNCNN model on a machine with an Intel i7-10700F CPU, an NVIDIA RTX 2070s GPU, and 32GB RAM.
Second, we record the enhancement time, PSNR before enhancement, and PSNR after enhancement of all the chunks into a dataset.
Third, we scale the enhancement time with four scale factors to model the different computational capacities. The four factors are} ultra-high computing capability (x4.5), high computing capability (x5), medium computing capability (x6), and low computing capability (x6.8), respectively.

\subsubsection{Metrics}
We select  {three sets of parameters for ($\alpha_1,\alpha_2,\alpha_3$) to indicate three QoE objectives, namely ($1,1,30$), ($1,1,60$), ($1,1,90$), for various latency tolerance.} 

\subsection{Baseline Schemes}
We compare the performance of ENAVS with the following:
\begin{itemize}
    \item \textbf{Bandwidth-based DASH (B-DASH):} The B-DASH policy takes the average download bandwidth of the past few chunks as the prediction of bandwidth,  {which is  five chunks in our simulation.}
    \item \textbf{Greedy algorithm (GA):} The GA takes the enhancement process into consideration with both download buffer and playback buffer.  {In our simulation, the GA} uses a greedy strategy to make the decision based on the maximal PSNR of the current chunk.
    \item \textbf{Pensieve:} Pensieve is an adaptive video streaming algorithm that based on A3C network to learn a streaming strategy without enhancement.
\end{itemize}

\subsection{Performance Analysis}
\subsubsection{QoE Performance}
\begin{figure}[!t] 
\centering
\includegraphics[width=0.8\linewidth]{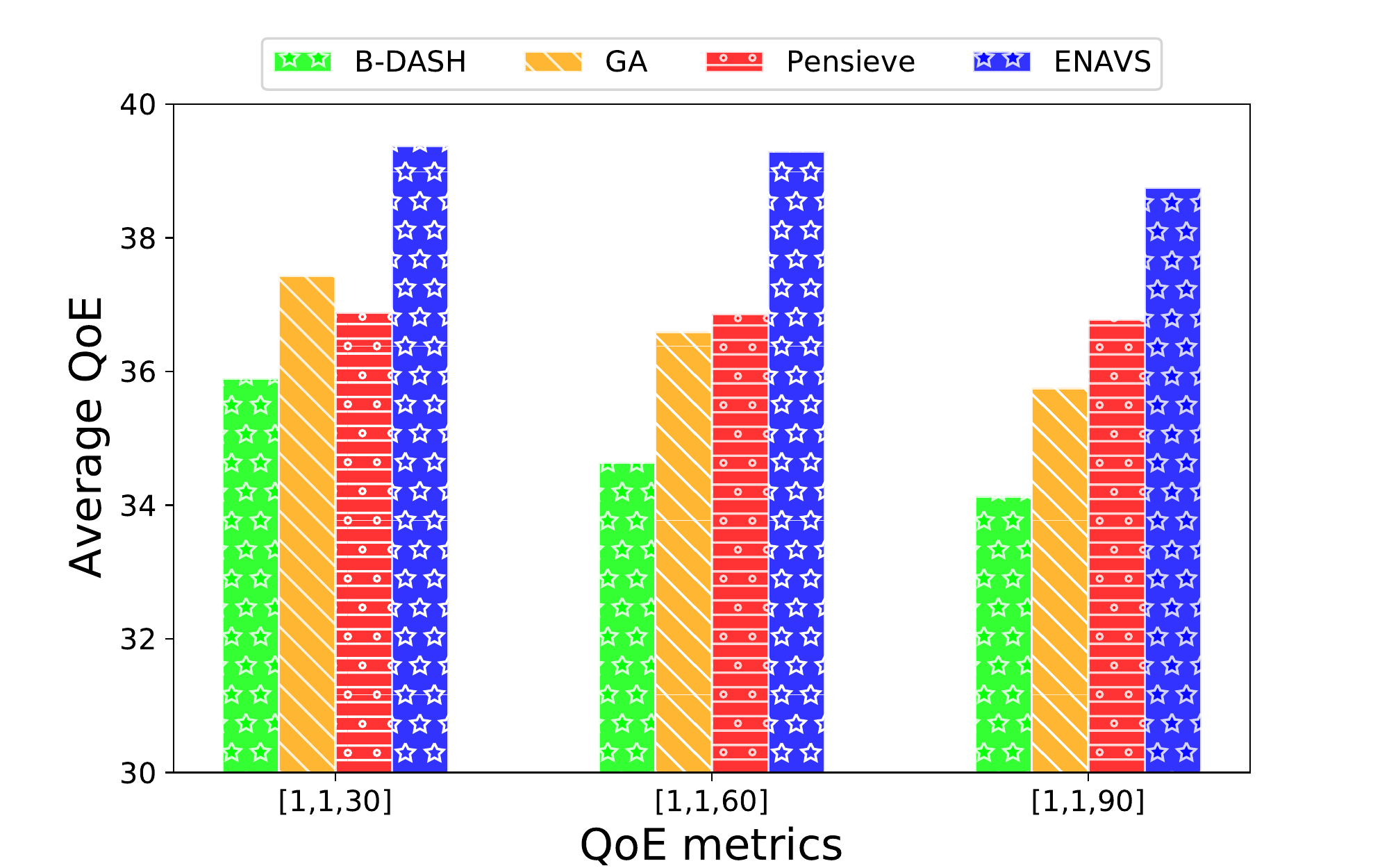}
\caption{Comparison of QoE under various QoE objectives}\label{qoe}
\end{figure}
In the first experiment, we evaluate the performance of ENAVS and the compared streaming system under various QoE objectives in Fig.~\ref{qoe}.
 {Each point is an average performance of 20 random seeds.}
We observe from the Pensieve and ENAVS in Fig.~\ref{qoe} that  the enhancement can increase the average QoE performance for all QoE objectives.
Moreover, ENAVS outperforms the  three baselines algorithm on all three QoE objectives with $5\% \sim 14\%$ improvement.
 {The major findings from the experimental
results can be summarized as follows:}
\begin{itemize}
    \item \textbf{B-DASH vs. Pensieve:} Pensieve improves the average QoE by $3\% \sim 8\%$ over B-DASH. The result shows that the RL-based adaptive video streaming model predicts the dynamic bandwidth more precise and selects a more suitable  bitrate to provide users with a better experience.
    \item \textbf{Pensieve vs. ENVAS:} The ENAVS improves the average QoE by $5\% \sim 7\%$ over Pensieve. The result shows that the enhancement module in the client can improve QoE of the downloaded chunks by enhancing the quality of them selectively.
     \item \textbf{GA vs. ENVAS:} The ENAVS improves the average QoE by $5\% \sim 9\%$ over GA, which indicates that the trained policy of the RL can adjust the bitrate of the download chunks and enhance factor dynamically to create a higher QoE than the greedy algorithm.
\end{itemize}


\subsubsection{Video Quality}
\begin{figure}
\begin{subfigure}{.49\linewidth}
\centering
\includegraphics[width=\linewidth]{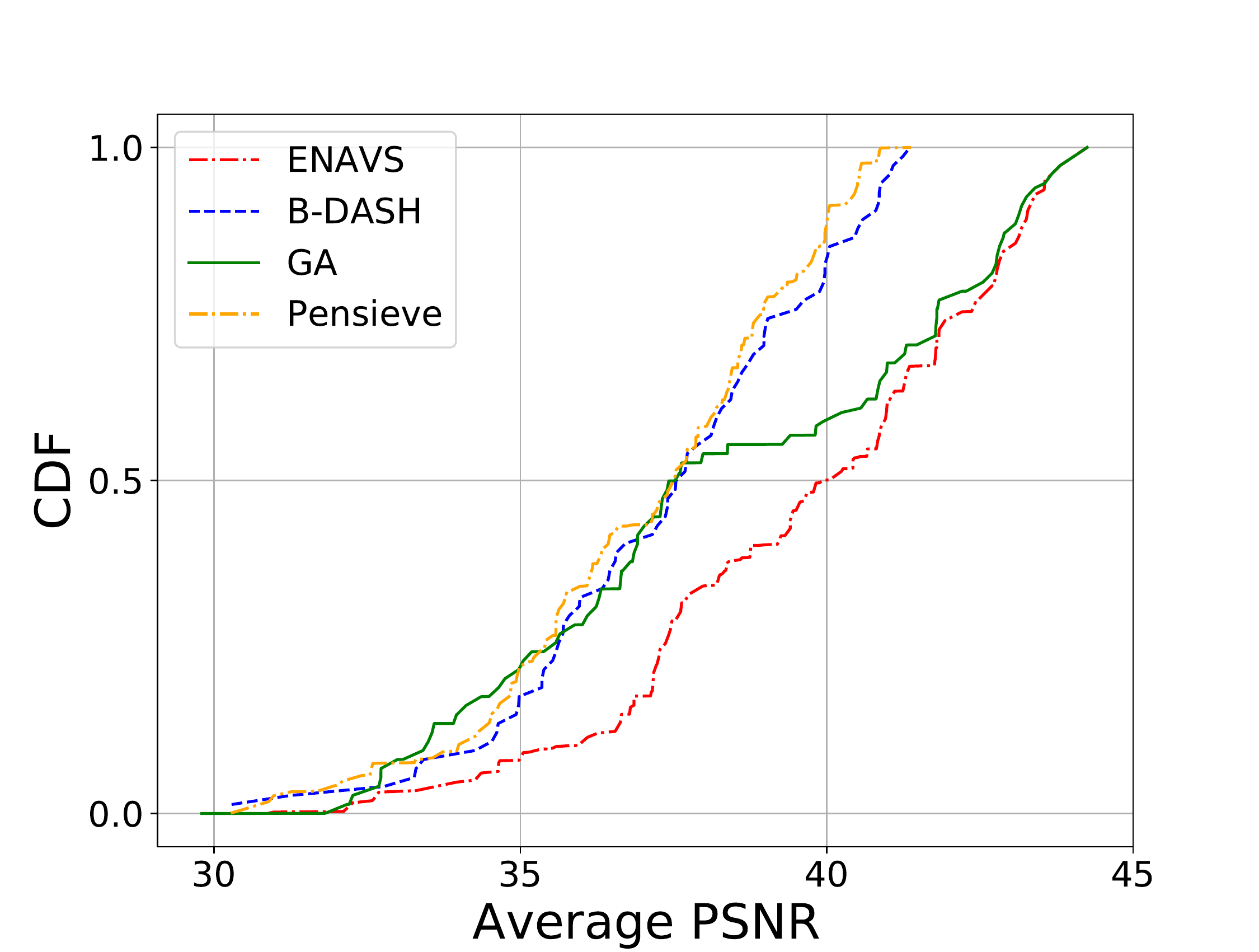}
\caption{High computing with (1,1,30)}
\end{subfigure}
\begin{subfigure}{.49\linewidth}
\centering
\includegraphics[width=\linewidth]{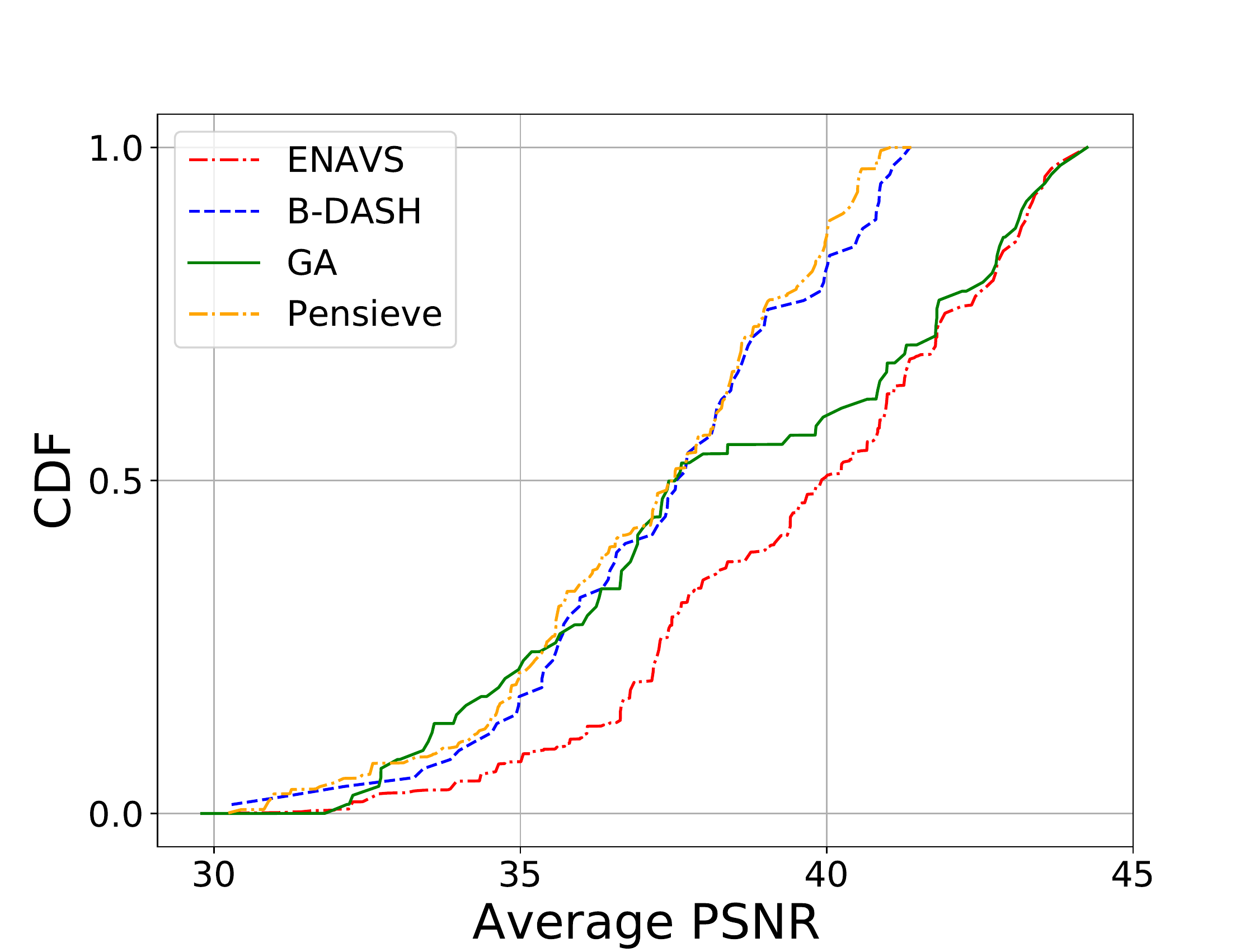}
\caption{High computing with (1,1,60)}
\end{subfigure}
\begin{subfigure}{.49\linewidth}
\centering
\includegraphics[width=\linewidth]{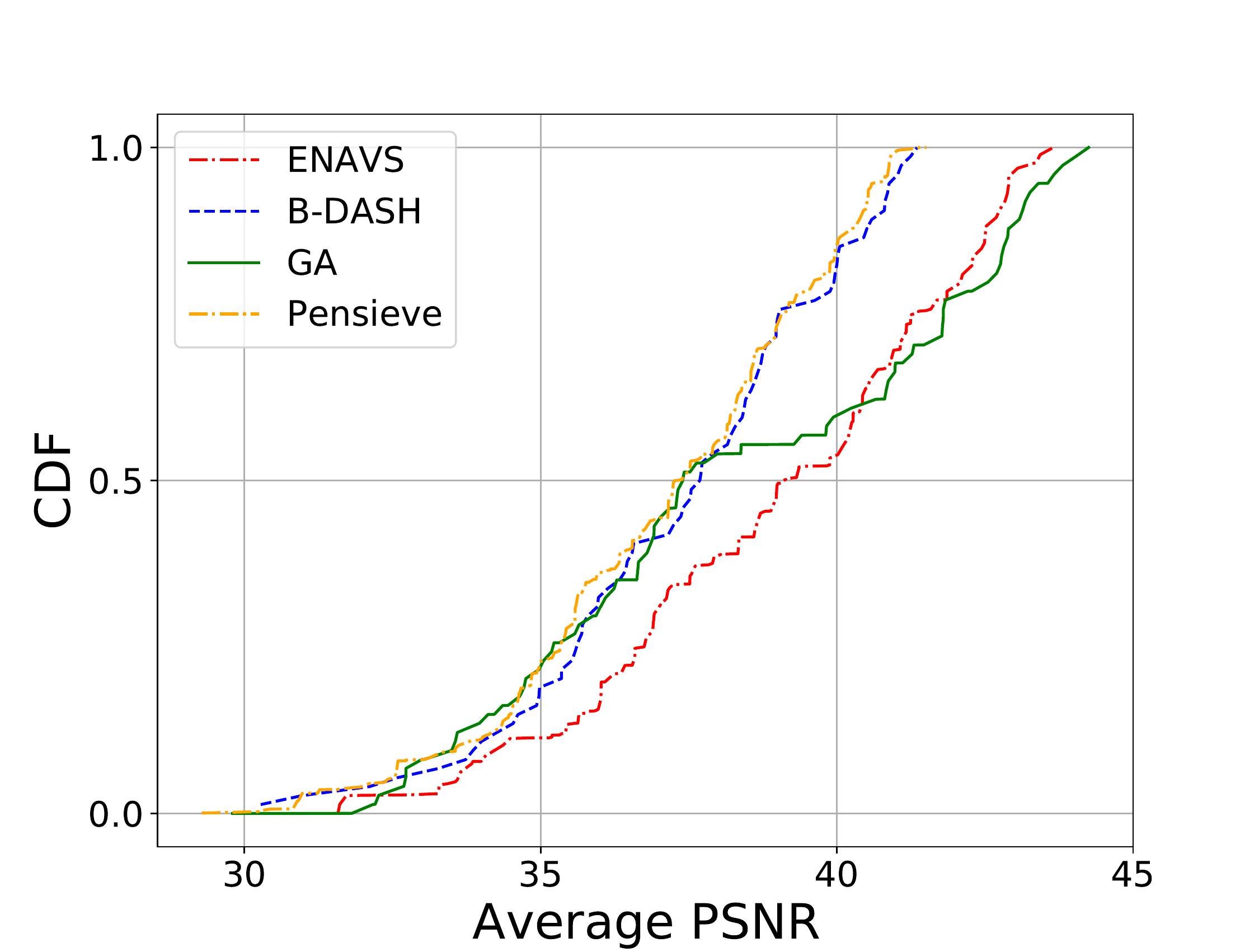}
\caption{High computing with (1,1,90)}
\end{subfigure}
\begin{subfigure}{.49\linewidth}
\centering
\includegraphics[width=\linewidth]{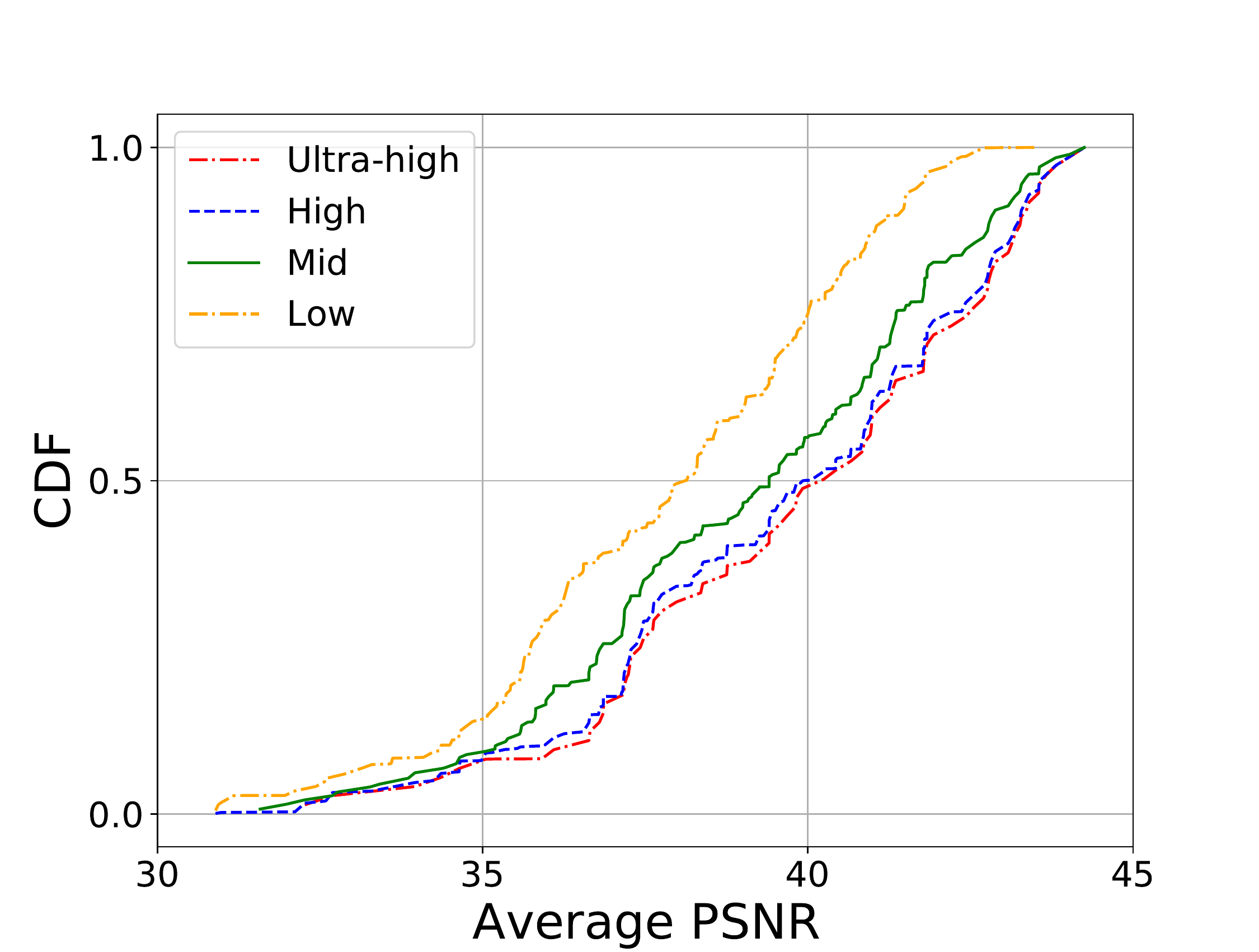}
\caption{(1,1,30)}
\end{subfigure}
\caption{Comparison of average PSNR versus different computing capabilities and $\bm{\alpha}$ }\label{fig:preference}
\end{figure}
 {In the second experiment,
we investigate the average PSNR distribution over different chunks, to illustrate how ENAVS increases the average QoE performance.
In particular, in Fig.~\ref{fig:preference}a-c, we plot the cumulative distribution functions (CDFs) of PSNR} that each streaming system achieves on the entire test data under three QoE objectives, respectively.
When the network condition is good, the effect of the GA is close to ENAVS; while in the poor network condition, the GA will degenerate to be similar to Pensieve.
The comparison indicates that the ENAVS provides video chunks with higher frame quality under the same network condition and computational capacity.
This is because the enhancement module in the system of ENAVS can improve the quality of the downloaded video chunks. 

\subsubsection{Average PSNR versus Computing Capability}

 {In the third experiment, we further compare the PSNR performance over different computational capabilities.
In particular, we plot the result under the QoE objective with $(1,1,30)$ in Fig.~\ref{fig:preference}d.
We observe similarly results under $(1,1,60)$ and $(1,1,90)$ and thus omit the results for page limit.}
We can observe from Fig.~\ref{fig:preference}d that devices with different computing capabilities present different patterns on the average PSNR of the video chunks.
Compared with the client with low computing power, the video quality played by the client with high computing power will be higher.
 {This validates that \emph{the proposed ENVAS adapts the policy to the client computation capability and thus could fully utilize the computation power from the heterogeneous client side.}}
Moreover, we find that the ultra-high computing capability device perceive similar video quality with the high computing capability device.
This observation verifies that ENAVS is practical to be implemented to the off-the-shelf playback devices.

\section{Conclusion}

In this paper, we have  integrated the video enhancement technique with the adaptive video streaming system.
We formulated an MDP problem that maximizes the QoE and proposed a DRL-based framework, called ENAVS.
ENAVS utilizes a DNCNN to improve the quality of low-bitrate video chunks downloaded over a poor wireless network.
Moreover, the learning-based ABR video streaming policy can adapt to the dynamic wireless network condition and the heterogeneous computing capabilities of the mobile client.
 The  simulation  shows  that  ENAVS  is  capable  of delivering $5\%\sim14\%$ more QoE under the same level of bandwidth and  computing  power  as  conventional  ABR  streaming.

\balance

\bibliographystyle{ieeetr}
\bibliography{ref.bib}

\end{document}